\documentclass[twocolumn,preprintnumbers,aps,superscriptaddress]{revtex4}

\def\bea{\begin{eqnarray}}
\def\eea{\end{eqnarray}}
\def\ben{\begin{equation}}
\def\een{\end{equation}}
\def\benu{\begin{enumerate}}
\def\enu{\end{enumerate}}

\def\sss{\scriptscriptstyle\rm}

\def\1var{(\bx_1...\bx\N)}

\def\br{{\bf r}}

\def\bx{{x}}

\def\x{_{\sss X}}

\def\s{_{\sss S}}
\def\xc{_{\sss XC}}
\def\Hxc{_{\sss HXC}}

\def\N{_{\sss N}}
\def\H{_{\sss H}}

\def\ext{_{\rm ext}}

\def\ee{_{\rm ee}}

\def\sph_int{ {\int d^3 r}}

\usepackage{graphicx}
\usepackage{psfrag}
\usepackage{amsmath}
\usepackage{bm}

\usepackage{epsfig}

\input epsf

\begin{document}
\title{Autoionizing Resonances in Time-Dependent Density Functional Theory}
\author{August J. Krueger}
\affiliation{Department of Physics and Astronomy, Hunter College and City University of New York, 695 Park Avenue, New York, NY 10065, USA\cite{1}}
\author{Neepa T. Maitra}
\affiliation{Department of Physics and Astronomy, Hunter College and The Graduate Center of the City University of New York, 695 Park Avenue, New York, NY 10065, USA}
\email{nmaitra@hunter.cuny.edu}
\date{\today}
\begin{abstract}
  Autoionizing resonances that arise from the interaction of a bound
  single-excitation with the continuum can be accurately captured with
  the presently used approximations in time-dependent density
  functional theory (TDDFT), but those arising from a bound double
  excitation cannot.  In the former case, we explain how an adiabatic
  kernel, which has no frequency-dependence, can yet generate the
  strongly frequency-dependent resonant structures in the interacting
  response function, not present in the Kohn-Sham response function.
  In the case of the bound double-excitation, we explain that a
  strongly frequency-dependent kernel is needed, and derive one as an {\it a
    posteriori} correction to the usual adiabatic approximations in
  TDDFT. Our approximation becomes exact for an isolated resonance in
  the limit of weak interaction, where one discrete state interacts
  with one continuum. We derive a ``Fano TDDFT kernel'' that
  reproduces the Fano lineshape within the TDDFT formalism, and also a
  dressed kernel, that operates on top of an adiabatic approximation.
  We illustrate our results on a simple model system.
\end{abstract}

\maketitle

\section{Introduction}
Since the birth of quantum mechanics, the study of photoionization 
has been important tool in characterizing the electronic structure of
materials. It is desirable for theoretical methods to supplement,
support, interpret, and even predict the experimental spectrum.
Resonance structures, arising from the interplay of bound and
continuum excitations (where this classification refers to some zeroth
order model), create a fascinating panorama of peaks in the spectrum,
whose profiles contain much information about the electronic states.
There are three major
issues to be surmounted in the theoretical treatment of resonances: first, resonances
require an adequate treatment of electron correlation.  We shall return to this point shortly, but note first that the
considerable advances in electronic structure methods and codes for
excitations over the years are predominantly set up for bound
states, not continuum states. Herein lies the second issue which is
adapting the many-body methods for non-square-integrable
scattering-type states~\cite{FB09,CCRM91}. Both basis set issues as well as the finite
matrix-based algorithms established in quantum chemistry codes need to
be revisited. When the system of interest is a molecule rather than an atom, 
a third ingredient compounds the problem: treating the correlated continuum states in multi-center non-spherical potentials. 
A variety of theoretical methods have been developed to treat these
issues to a variety of extents, and we mention only a smattering of
these here. For atoms, one of the more successful approaches is
multi-configuration Hartree-Fock~\cite{F78}; this accounts for
electron correlation and core polarization accurately
enough to describe atomic resonances eg. in halogens~\cite{S06}. For molecules, Ref.~\cite{SGWF89} studied the relationship between interatomic distances and resonance positions using a minimal-basis static exchange method. Another approach utilizes R-matrix
theory within multi-channel quantum defect theory, eg.  for the calcium atom in
Ref.~\cite{KG88}. Reformulating the scattering problem as a bound-state problem in this way means that advanced electronic structure codes may be used. A complex-scaled
full-configuration-interaction method was used in Ref.~\cite{SM08} to
calculate the resonances in a two-electron quantum dot.
Methods using a complex absorbing potential in a configuration-interaction calculation, or with  a correlated independent particle potential have been developed;   an inner-valence autoionizing resonance of the neon-dimer Ne$_2^+$~\cite{SC01}, the nitrogen dimer N$_2$ and acetylene C$_2$H$_2$~\cite{SSP05}, for example, were computed in this way. Several other works studied autoionizing resonances in acetylene, e.g. Ref.~\cite{LL00} used configuration interaction within the multichannel Schwinger variational method, and Ref.~\cite{CMR00} used $L^2$ Gaussian-type orbital basis sets. 

Accounting for electron correlation becomes an increasingly Herculean task for wavefunction-based methods as the number of electrons in the system grows.
Time-dependent density functional theory (TDDFT) ~\cite{RG84,GDP96,TDDFTbook} is a particularly
efficient approach to the many-body problem, making it attractive
for calculating photo-ionization spectra of chemically interesting
systems.  Recent work by Stener, Decleva, Fronzoni, and co-workers,
Refs.~\cite{SFD05,FSD03,FSD04,STFD06,STFD07,SDL95} (and references therein),
has shown that TDDFT predicts accurate resonance parameters for a
range of medium-size molecules, including acetylene, carbon monoxide, silicon tetrafluoride, and sulphur hexafluoride. Their earlier works used a one-center expansion B-spline basis, while the later ones
utilized multi-centric B-spline basis functions, more suited to larger
molecules.

Due to its favorable system-size scaling, TDDFT has become a method of
choice for the calculation of bound spectra in quantum chemistry. For example, 
it allows calculations of spectra of systems as large as
biomolecules (see eg. Refs.~\cite{FA02,SRVB09,MLVC03}), and
coupled electron-ion dynamics on non-trivial chemical
reactions~\cite{TTRF08}.  One first computes the Kohn-Sham (KS)
spectrum, which is the response of the non-interacting KS system: the
single-particle excitations and oscillator strengths of the
ground-state KS potential. The  true spectrum 
is obtained by applying the linear response TDDFT exchange-correlation (xc)
kernel (see also Sec.~\ref{sec:ATDDFT}): operating via a matrix
equation or a Dyson-type integral equation, the kernel mixes the
single-excitations of the KS system, and, were the exact kernel known,
this would yield the exact spectrum of the true interacting system. In
practise, approximations are needed for the kernel, as well as for the
ground-state KS potential out of which the KS spectrum is calculated.
Of particular note for the present paper, is that almost all
calculations use an adiabatic approximation to the kernel, meaning one
that has no frequency-dependence.

In the TDDFT calculations of auto-ionization cited above, the role
of the channel coupling is very clear: the bare KS spectra are smooth and
relatively featureless, while after the TDDFT procedure is applied, resonances are generated. 
In one of these earliest TDDFT calculations
(in the Ne atom~\cite{SDL95}), it was noted that while the resonances
arising from bound single KS excitations whose energy lie in the
continuum (eg. core to Rydberg excitations) are quite accurately
predicted, those arising from bound double-excitations, are totally
missed. (This was also noted to occur in acetylene~\cite{FSD04}). This was explained in Ref.~\cite{SDL95} as arising because the linear response method involves only
first-order changes in the density (or wavefunction), and therefore
only single-excitations can be obtained. However, in principle, TDDFT
linear response reproduces {\it all} excitations of the system, which
may be linear combinations of Slater determinants with any number of
excited electrons. The lack of resonances from double-excitations is {\it not} a failing of TDDFT, but rather is a failing of the approximation for the xc kernel that is used.
For {\it bound}
double-excitations, Ref.~\cite{MZCB04} showed that 
the exact kernel is necessarily
strongly-frequency-dependent in the neighbourhood of a state of
double-excitation character. An approximate
frequency-dependent kernel was derived there to account for
double-excitations; this was successfully tested on real molecules in
Refs.~\cite{CZMB04,MW08}.

In the present paper, we investigate the form of the kernel that is
needed in order to capture autoionizing resonances arising from a
bound double-excitation. 
Our derivation essentially adapts Fano's 1961
analysis~\cite{F61} to the case when the ``unperturbed'' states in his
configuration mixing are the relevant KS bound-state and the continuum
its energy lies in. We begin therefore, in
Sec.~\ref{sec:Fano}, with a brief recapitulation of Fano's formula.
In section~\ref{sec:ATDDFT}, we discuss the implications of Fano's
formula for the density-density response functions of TDDFT,
explaining with an illustrative sketch, that while a strong
frequency-dependence is required in the TDDFT kernel to capture
resonances arising from a bound double-excitation, resonances arising
from bound single excitations are captured by the usual adiabatic
(non-frequency-dependent) TDDFT kernels.  Then in
section~\ref{sec:dressed}, we derive an approximation, in the spirit
of Fano, which does capture the double-excitation resonance, and
illustrate it on a simple model system.  Our approximation becomes
exact in the limit of weak interaction, for an isolated narrow resonance in a single continuum.

\section{Fano Resonances}
\label{sec:Fano}
 The universality of the Fano profile has been noted by
 many, from lineshapes in spectra of atoms, molecules, solids and heterostructures, to
 interference in quantum dots and Aharanov-Bohm rings; its
 robustness reflected in the more than 3000 citations of his 1961
 paper, Ref.~\cite{F61}.   The simplest type
 of resonance occurs when a single bound-state interacts with a single
 continuum.  In the absence of this interaction, the continuum is assumed to be relatively ``flat'', i.e. featureless. 
Fano performed a careful diagonalization of the
 Hamiltonian for a continuum coupled to a single bound state $\vert \Phi_{\rm b}\rangle$ whose
 energy lies in the continuum~\cite{F61}. 
 Denoting the coupling Hamiltonian by $V_{\rm cpl}$, one defines the matrix element, $V_{E} = \langle\Phi_{E}\vert V_{\rm cpl}\vert \Phi_{\rm b}\rangle$, with $\Phi_E$ being the (uncoupled) continuum state. 
Fano derived the following formula for
 the matrix element of some transition operator $\hat{T}$ (eg. a
 dipole operator) between an initial (bound) state $\vert i\rangle$ and a
 state $\Psi_E$ lying in the resonance region, resulting from the diagonalization:
\ben
\frac{\vert\langle\Psi_E\vert \hat{T}\vert i\rangle\vert^2}{\vert\langle\Phi_E\vert \hat{T}\vert i\rangle\vert^2} = \frac{(\omega - \omega_r + \Gamma q/2)^2}{(\omega - \omega_r)^2 + (\Gamma/2)^2}
\label{eq:fano}
\een
Atomic units are used throughout this paper. 
Here $\omega = E - E_i$ is the frequency, and
\ben
\omega_r = E_r -E_i = E_{\rm b} - E_i + P\int\frac{\vert V_{E'}\vert^2}{E-E'}dE'
\label{eq:wr}
\een
is the ``position'' of the resonance, shifting the unperturbed bound-state frequency $E_{\rm b} - E_i$ by the principal-value integral. The parameter
\ben
\Gamma = 2\pi\vert V_E\vert^2
\label{eq:gamma}
\een
defines the width of the resonance, while the parameter $q$ characterizes its asymmetry:
\ben 
q = \frac{\langle\Phi_{\rm b}\vert\hat{T}\vert i \rangle + P\int V_{E'}\langle\Phi_{E'}\vert\hat{T}\vert i \rangle/(E-E') dE' }{\pi V_E\langle\Phi_E\vert\hat{T}\vert i \rangle}
\label{eq:q}
\een 
For example, $q=0$ represents a negative purely symmetric
Lorentzian, $q\to\infty$ represents a positive purely symmetric
Lorentzian, and $q = \pm 1$ represents a purely antisymmetric
lineshape. The asymmetry can be interpreted as a consequence of interference between the autoionizing state and the continuum states~\cite{F35,F61,FC65}. In Refs.~\cite{F61,FC65}, it is argued that typically $q$ is negative.

Although Eqs.~(\ref{eq:gamma}) and~(\ref{eq:q}) appear energy-dependent, $q$, $\Gamma$, and $\omega_r$
are regarded as constant through the resonance region. For a narrow enough
resonance, this is a reasonable assumption; essentially the idea is that
$\Gamma$ is the smallest energy scale in the system. 
Fano also derived a ``sum-rule'' for the integrated transition probability:
\bea
\nonumber
\vert\langle\Phi_{\rm b}\vert T\vert i\rangle\vert^2&=&
\int dE\left(\vert\langle\Psi_E\vert T\vert i\rangle\vert^2 -\vert\langle\Phi_E\vert T\vert i\rangle\vert^2\right) \\
&=& \vert\langle\Phi_E\vert T\vert i\rangle\vert^2\frac{\pi}{2}(q^2-1)\Gamma 
\label{eq:sumrule}
\eea
which expresses the unitary nature of the diagonalization procedure. Eq.~\ref{eq:sumrule} is essentially a consequence of the following closure relation:
\ben
\int \vert \Psi_E\rangle\langle\Psi_E\vert dE = \int \vert \Phi_E\rangle\langle\Phi_E\vert dE + \vert \Phi_{\rm b}\rangle \langle\Phi_{\rm b}\vert 
\label{eq:closure}
\een

Fano's Eq.~(\ref{eq:fano}) tells us that the transition to the
continuum when a discrete state couples to the continuum, is equal to
that without the coupling, multiplied by a (generally asymmetric)
Lorentzian line-shape factor.

Photoabsorption cross-sections measure the dipole transition
probability, so  take $\hat T$ to be the dipole operator.
Fits are routinely made for the Fano parameters $q,\Gamma$ and
$\omega_r$ for a given cross-section obtained from experiment or
theory, i.e. Eqs.~(\ref{eq:wr} --~\ref{eq:q}) are not typically 
used to calculate these quantities, rather, they are extracted from 
experimental or theoretical data.   Although we will only use the simplest 
type of resonance in the present paper, we do note that Fano's analysis has been 
generalized in several directions~\cite{F61,FC65,S77},  e.g. when more continua are present.
The more complicated situations involve more fitting parameters, and in practise fits are made for these more general formulae, rather than the simplest situation discussed above.

\section{Autoionizing Resonances within Adiabatic TDDFT}
\label{sec:ATDDFT}

\subsection{Photoabsorption/ionization in TDDFT}
The photoabsorption or photoionization cross-section measures the linear density response $\delta n(\br,\omega)$ of a system subject to an external electric field of frequency $\omega$.
Letting $\alpha, \beta$ denote the three spatial directions, 
the cross-section tensor is
\ben
\sigma_{\alpha\beta}(\omega) =\frac{4\pi\omega}{c}\int d^3r d^3r' r_\alpha r'_\beta Im\chi(\br,\br',\omega)
\label{eq:crosssection}
\een
where $\chi(\br,\br',t-t') = \delta n(\br,t)/\delta v\ext(\br',t)$ is
the density-density response function of a system to an external potential $v\ext(\br,t)$ (see, eg. Refs.~\cite{Friedrich,GVbook}). 
A sum-over-states expression may be obtained using the standard linear response theory expansion for the density-density response function:
\ben
\chi(\br,\br',\omega) = \sum_E\frac{\langle 0\vert \hat{n}(\br)\vert E\rangle\langle E\vert \hat{n}(\br')\vert 0\rangle}{\omega - (E - E_0) + i0^+} - \frac{\langle 0\vert \hat{n}(\br')\vert E\rangle\langle E\vert \hat{n}(\br)\vert 0\rangle}{\omega + (E - E_0) + i0^+}
\label{eq:chi_sos}
\een
where $\vert E\rangle$ label the excited states, and $\vert 0\rangle$ is the ground-state with energy
$E_0$. $\hat{n}(\br) = \sum_{i}^N\delta(\br - \br_i)$ is the density
operator. 
Assuming real eigenstates, the imaginary part is extracted as
\bea
\nonumber
\Im\chi(\br,\br',\omega)&=& -\pi\sum_E\langle 0\vert \hat{n}(\br)\vert E\rangle\langle E\vert \hat{n}(\br')\vert 0\rangle\times\\
&&\left(\delta(\omega - \omega_E) - \delta(\omega + \omega_E)\right)
\label{eq:imchi_gen}
\eea
where $\omega_E= E-E_0$.
Inserting this into Eq.~\ref{eq:crosssection}, we have
\bea
\nonumber
\sigma_{\alpha\beta}(\omega) &=& -\frac{4\pi^2\omega}{c}\sum_E\langle 0\vert \hat\br_\alpha\vert E\rangle\langle E\vert \hat\br_\beta'\vert 0\rangle\times\\
\nonumber
&&\left(\delta(\omega - \omega_E) - \delta(\omega + \omega_E)\right)\\
&=& -\frac{4\pi^2\omega}{c}\sum_E d_\alpha(\omega)d_\beta(\omega)\left(\delta(\omega - \omega_E) - \delta(\omega + \omega_E)\right)
\label{eq:sigma_sos}
\eea
where $d_\alpha(\omega)$ is the transition dipole moment from the ground-state to the excited state of energy $E = \omega +E_0$. 

Therefore, to compute photoabsorption or photoionization, one needs an efficient way to calculate
either the density response $\delta n(\br,\omega)$, or the
density-density response function, $\chi(\br,\br',\omega)$ (Eq.~\ref{eq:crosssection}) or the
excited states of the system (Eq.~\ref{eq:sigma_sos}). Given that the electrons in the system
are interacting with all the others via Coulomb repulsion, this becomes a
daunting task for correlated wavefunction methods for all but the smallest molecules: the numerical effort in
solving the problem scales exponentially with the number of electrons
in the system. 

In TDDFT, instead of dealing with the correlated many-body wavefunction, one solves 
for  the much simpler single-particle orbitals $\phi_i(\br,t)$ that evolve in the one-body KS Hamiltonian.
Nevertheless, from the $\phi_i(\br,t)$
in principle, not just the exact density but 
{\it all} properties of the true interacting correlated system may be extracted~\cite{RG84,TDDFTbook}. In practise approximations are needed for the xc terms.

Linear response TDDFT~\cite{PGG96,C96} is founded on the
relation between the interacting and KS density-response functions, both of which are functionals of the ground-state density of the system of interest, $n_0(\br)$.
First, we note that for non-interacting systems such as KS, the numerator in Eq.~\ref{eq:chi_sos} simplifies to products of occupied (indexed by $i$) and unoccupied (indexed by $a$) orbitals:
\ben
\chi\s(\br,\br',\omega)=\sum_{i,a}\frac{\Phi_{ia}(\br)\Phi_{ia}^*(\br')}{\omega - (\epsilon_a - \epsilon_i) + i0^+}- \frac{\Phi_{ia}^*(\br)\Phi_{ia}(\br')}{\omega + (\epsilon_a - \epsilon_i) + i0^+}
\label{eq:chis_sos}
\een
where $\Phi_{ia}(\br) = \phi_i^*(\br)\phi_a(\br)$. 
The orbital energy-differences in the denominator $\epsilon_a - \epsilon_i$, are the KS single-excitation frequencies, 
and the orbitals live in the KS potential $v\s(\br) = v\ext(\br) + v\H(\br) + v\xc(\br)$. Here $v\H(\br) = \int d^3r' n(\br')/\vert \br - \br'\vert$ is the Hartree potential, and $v\xc(\br)$ is the (ground-state) xc potential. 
The frequencies of the true system lie at the poles of the true interacting response function, $\chi(\br,\br', \omega)$:
\bea
\nonumber
\chi[n_0](\br,\br',\omega) = \chi\s[n_0](\br,\br',\omega) + \\
\int dr_1 dr_2 \chi\s[n_0](\br,\br_1,\omega)f\Hxc[n_0](\br_1,\br_2,\omega)\chi[n_0](\br_2,\br',\omega)
\label{eq:dyson}
\eea
where $f\Hxc$ denotes the ``Hartree-exchange-correlation kernel'', 
\bea
\nonumber
f\Hxc[n_0](\br,\br',\omega) \equiv \frac{1}{\vert \br -\br'\vert} + f\xc[n_0](\br,\br'\omega)\\
=\chi\s^{-1}[n_0](\br,\br',\omega) - \chi^{-1}[n_0](\br,\br',\omega)
\label{eq:fHxc}
\eea
with $f\xc[n_0](\br,\br',t-t') = \delta v\xc[n_0](\br,t)/\delta n(\br',t)$. 


Although the exact xc kernel, $f\xc(\omega)$ is frequency-dependent,
reflecting a dependence on the history of the density in the
time-domain~\cite{MBW02}, the majority of calculations
today utilize an ``adiabatic'' approximation, meaning one where
$v\xc[n](r,t)$ depends only on the instantaneous density
$f\xc^{A}[n_0](\br,\br',t-t') \propto \delta(t-t')$.  When
Fourier-transformed, this gives no structure in frequency-space.
Instead, the adiabatic approximation is based on a ground-state energy
functional 
\ben
f\xc^{A}[n_0](\br,\br') = \left.\frac{\delta^2 E\xc[n]}{\delta n(\br')\delta n(\br)}\right\vert_{n=n_0}
\label{eq:fHxcA}
\een

Despite incorrectly lacking frequency-dependence, the adiabatic
approximation yields remarkably accurate results for most excitations.
Eq.~\ref{eq:dyson} can be transformed to a matrix
equation indexed by KS single-excitations~\cite{C96}, the workhorse of bound-state TDDFT.
When the true interacting state is composed of mixtures of KS
single excitations, adiabatic TDDFT is expected to work reasonably
well -- provided the spatial functional dependence of $v\xc(\br)$
is adequately nonlocal for the problem at hand. 
For states of multiple-excitation
character however, one must go beyond the adiabatic approximation as
was shown for bound states of finite
systems~\cite{MZCB04,CZMB04,C05,MW08,TH00}, as mentioned in the
introduction. We shall show shortly, that when such a state lies
in the continuum, a frequency-independent kernel misses the
resonance arising from it.

\subsection{Autoionizing Resonances in TDDFT}
In photo-ionization, the energy of interest lies in the continuum; the sum over the delta-function peaks in Eqs.~(\ref{eq:imchi_gen}) and~(\ref{eq:sigma_sos}) becomes an integral, and, (for positive $\omega$, greater than the ionization threshold),
\ben
\Im\chi(\br,\br',\omega) = -\pi\left\langle 0\vert \hat{n}(\br)\vert E=\omega+E_0\right\rangle\langle E=\omega+E_0\vert \hat{n}(\br')\vert 0\rangle
\label{eq:imchi_ctm}
\een
where the continuum states are chosen  real and energy-normalized. 
The KS cross-section for continuum
transitions are largely structureless, especially when only one continuum is relevant, the KS transition dipole is:
\ben
 {\bf d}\s(\omega) = \langle \phi_{\epsilon= \epsilon_i+\omega} \vert \br\vert\phi_i\rangle 
\label{eq:KSdipole}
\een
where $\phi_i$ is the KS occupied orbital, of energy $\epsilon_i$, out of which excitation occurs to the continuum orbital $\phi_\epsilon$, of energy $\epsilon = \epsilon_i+\omega$. 
This typically gently decays as a function of
frequency, reflecting the decay of the overlap between an occupied
orbital and a continuum one as the energy of the continuum state
rises.  

When one applies TDDFT to obtain the spectrum of the
interacting system, the kernel $f\Hxc$ mixes the KS single
excitations and the spectrum distorts to varying degrees:
less at the higher frequencies, and most significantly near
resonances. The TDDFT kernel smears the oscillator strength from
bound transitions whose energy lies in the continuum over a narrow
range in the continuum, implicitly performing the job of the Fano
diagonalization of Section~\ref{sec:Fano}. Effectively, a rather
featureless KS continuum spectrum is turned into a dramatically
frequency-dependent interacting spectrum, via the operation
of the xc kernel.
Refs.~\cite{SFD05,FSD03,FSD04,STFD06,STFD07,SDL95} have demonstrated this explicitly
on a wide range of interesting atoms and molecules. These works show
that adiabatic kernels reproduce resonance features of interacting
systems rather well through this action, when the resonances arise
from a bound single excitation with energy lying in the continuum (e.g. of
core-Rydberg nature).  

The appearance of the resonance can therefore be understood
as a mixing of the single excitations appearing in Eq.~(\ref{eq:chis_sos}) via Eq.~(\ref{eq:dyson}).  
Yet, how a frequency-independent kernel can transform the largely
frequency-independent KS continuum spectrum into a spectrum that does
have such dramatic frequency-dependence, may strike one as
incongruous. To understand this better, we first find an expression relating the interacting and KS response functions near an autoionizing resonance, following in Fano's footsteps. 

We keep to the simple case of one discrete state lying in one
continuum. Further, we consider, initially, {\it only} the resonant
coupling; that is, we treat the KS states as Fano's
``pre-diagonalized'' states, without accounting for mixing amongst
them. Although this could only be the case when there is no electron-interaction, the justification for this simplification is that in the vicinity of the resonance, this resonant coupling is certainly the dominant effect: the coupling amongst the KS states is an order of $\Gamma$ less. (Later, in Sec~\ref{sec:dressed}, we relax this assumption).
For the matrix elements on the right-hand-side of
we use Eq.~(\ref{eq:fano}), but taking $\hat{T}$ as the density operator
$\hat{n}(\br)$:
\ben
\langle 0\vert \hat{n}(\br)\vert E\rangle = \sqrt{\frac{(\omega - \omega_r + \Gamma q(\br)/2)^2}{(\omega - \omega_r)^2 + (\Gamma/2)^2}}\langle 0\vert \hat{n}(\br)\vert E\rangle\s 
\label{eq:trans1}
\een
where 
the ket on the right is the KS
continuum excited state at energy $E = \omega +E_0$. 
The state $\vert 0\rangle$ on both sides of the equation is
 the initial state out of which transitions occur, which we will
take to be the ground-state. Technically, this should be the
interacting ground-state because Fano's analysis assumed everything except
for the resonant coupling was in the zero-order states. However, if
we approximate the bra $\langle 0\vert$ on the right of Eq.~(\ref{eq:trans1}) to be
instead the KS ground-state,  this formula then directly relates the
interacting matrix element (related to the oscillator strength) to the
KS matrix element.  This approximation holds if the KS
ground--continuum transition in the absence of the resonance is
a good approximation to the true interacting transition, and again, holds under the justification that the resonant coupling is the dominant effect in the resonance region. 


The quantities $\Gamma$, $q$ and $\omega_r$ are given by
Eq.~\ref{eq:gamma},~\ref{eq:q}, and~\ref{eq:wr}, where the
``coupling'' $V_{\rm cpl}$ is the difference in the Hamiltonian of the
true and KS systems:
\ben
V_{\rm cpl}  = V\ee - v\H - v\xc
\label{eq:Vcpl}
\een
where $V\ee = \frac{1}{2}\sum_{ij}1/\vert \br_i - \br_j\vert$ is the electron-electron interaction.
Inserting Eq.~\ref{eq:trans1} into Eq.~\ref{eq:imchi_ctm}, for $\omega$ near a resonance, 
\bea
\nonumber
\Im\chi(\br,\br',\omega) &=& \frac{(\omega - \omega_r + \Gamma q(\br)/2)(\omega - \omega_r + \Gamma q(\br')/2)}{(\omega - \omega_r)^2 + (\Gamma/2)^2}\\
&\times&\Im\chi\s^{\rm c}(\br,\br',\omega)
\label{eq:imchi}
\eea
where $\chi\s^{\rm c}(\br,\br',\omega)$ denotes the continuum contribution to the KS response function at frequency $\omega$. 
Following directly from Fano's
analysis, Eq.~\ref{eq:imchi} is a new relation between the true and KS
response functions in the neighborhood of any isolated narrow autoionizing resonance,
that arises from the mixing of a single discrete state with a single continuum, 
under the assumption of weak interaction. The real part of the response function can be obtained via the principle-value integral:
\ben
\Re\chi(\br,\br',\omega) =\frac{2}{\pi}P\int_0^{\infty} \frac{\omega'\Im\chi(\br,\br',\omega)}{\omega'^2 - \omega^2}d\omega'
\label{eq:real}
\een
due to known analyticity properties of $\chi$ (eg. Ref.~\cite{GVbook}).

\begin{figure}[h]
\centering
\includegraphics[height=4.5cm,width=8.5cm]{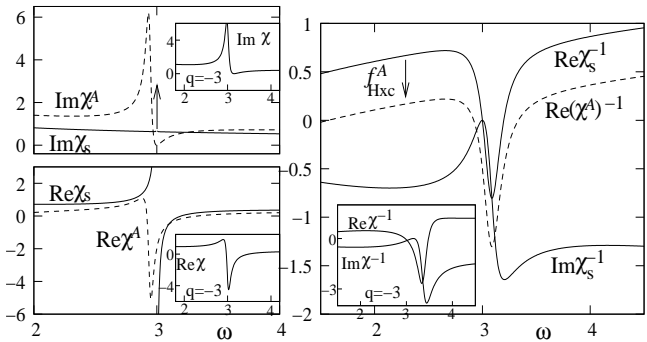}
\caption{The imaginary and real parts of $\chi\s(\omega)$ (left-hand panels, solid lines).   The delta-function in the imaginary part of $\chi\s$ corresponding to the bound-state, is shown as an arrow.  On the right, we show $\chi\s^{-1}$ (solid) and the shift that an adiabatic kernel $f\Hxc^{A}$ produces, yielding $\left(\chi^{A}\right)^{-1}$ (dashed). When inverted to yield the adiabatic response kernel  $\chi^{A}(\omega)$, the resonance emerges, as indicated by the dashed lines on the left panels. The adiabatic curves  should be compared with the  insets, which come from Eq.~\ref{eq:imchi} with $q=-3$, see text}
\label{fig:adia_sing}
\end{figure}

We now address the curiousity raised earlier regarding how application of a
frequency-independent adiabatic kernel to the flat KS spectrum can generate frequency-dependent resonant structure in
$\chi$.
A simple sketch is instructive to show this.  Consider a resonance due
to a bound single-excitation, at frequency $\omega_b>I$, the ionization potential. Within our weak-interaction assumption the KS single excitation
couples only to the continuum states,  so, near $\omega_{\rm b}$, $\chi\s$ has the form
\ben
\chi\s(\omega) = \chi\s^{\rm c}(\omega) + \chi\s^{\rm b}(\omega)
\label{eq:chis_sing}
\een
where $\chi\s^{\rm c}(\omega)$ is complex, smooth and gently-decaying and $\chi\s^{\rm b}(\omega)$ is the contribution of the bound state to the sum over states Eq.~(\ref{eq:chis_sos}).
For the present illustrative purposes, we neglect the spatial-dependence. 
We take $\chi\s^{\rm c}(\omega)
 =  \left(\frac{1}{\sqrt{\omega +I}} + \frac{i}{\sqrt{\omega -I}}\right)$ in the plots, although our conclusions in no way depend on this form,  and $\chi\s^{\rm b}(\omega)= \frac{b}{\omega - \omega_{\rm b}+ i0^+}$. Here $b$ represents the orbital products appearing in the residue of Eq.~(\ref{eq:chis_sos}). The imaginary part of this $\chi\s(\omega)$ is plotted as the solid line on the top left panel of Figure~\ref{fig:adia_sing}. 
In particular, note that $\Im\chi\s$, which is directly related to the measured cross-section (Eq.~\ref{eq:crosssection}), has the structureless continuum contribution plus a delta-peak (indicated by the arrow) at $\omega_{\rm b}$ (taken to be 3, while $b$ is taken to be -0.01 in these plots and $I=0.5$au); other values yield similar plots).  (Note that the delta-peak  is not evident in the smooth
KS cross-sections plotted in the graphs of
Refs.~\cite{SFD05,FSD03,FSD04,STFD06,STFD07,SDL95}, because there only the KS continuum transitions in
$\chi\s^{\rm c}$ (Eq.~(\ref{eq:KSdipole})) are included).
Now consider
inverting Eq.~\ref{eq:chis_sing}. Interestingly, this immediately
displays resonance-structures in both the real and imaginary part of
$\chi\s^{-1}(\omega)$, as shown by the solid lines on the right-hand panel of Figure~\ref{fig:adia_sing}. This may be simply seen mathematically: inverting Eq.~\ref{eq:chis_sing} reveals a Lorentizan denominator in both the real and imaginary parts. 
Applying the TDDFT kernel, to obtain $\chi^{-1} = \chi\s^{-1} - f\Hxc$
(Eq.~(\ref{eq:fHxc})), we see that an adiabatic approximation
$f\Hxc^{A}$ (Eq.~(\ref{eq:fHxcA})) just uniformly shifts the real part of
$\chi\s^{-1}$, while not adding any additional frequency-dependence. Indeed the resonance structure of $\chi\s^{-1}$ and the
adiabatically shifted one $(\chi^A)^{-1}$ resemble that of the Fano
profile of Eq.~(\ref{eq:imchi}):  for a constant $q=-3$, the inverse of $\chi$ computed from Eqs.~(\ref{eq:imchi}--\ref{eq:real}) is plotted in the inset on the right. No frequency-dependence is required in the kernel itself to
obtain resonances in $\chi\s$.  We now invert $(\chi^A)^{-1}$ to obtain $(\chi^A)$, plotted back as the dashed curves in the left-hand panels
of Figure~\ref{fig:adia_sing}. 
These resemble the Fano profile of  Eq.~(\ref{eq:imchi}) for $q=-3$ plotted in the insets on the left panels. 
 The adiabatic shift in the inverse response functions, had
the effect of Lorentzian smoothing the delta-function peak in
$\Im\chi\s$, producing a resonance in $\Im\chi^{A}$.  Therefore,
making a adiabatic shift in the real part of the inverse response fn,
$\Re\chi\s^{-1}$, turns, upon inversion, a delta-peak of $\Im\chi\s$
into a Lorentzian resonance peak of $\Im\chi^{A}$. In this way, {\it inversion of $\chi\s^{-1}$, shifted 
frequency-independently,  generates
frequency-dependent structure in $\chi$.}


\begin{figure}[h]
\centering
\includegraphics[height=4.5cm,width=8.5cm]{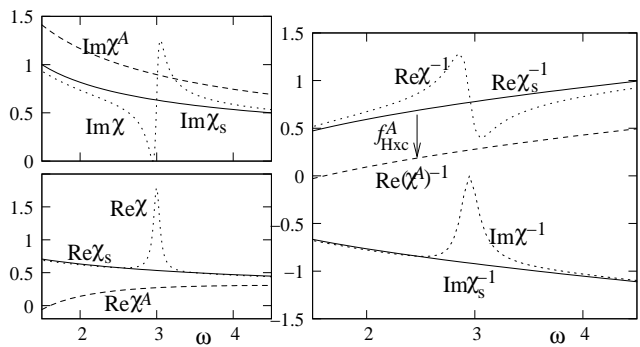}
\caption{The imaginary and real parts of $\chi\s(\omega)$ (left-hand panels, solid lines). 
On the right panel, we show $\chi\s^{-1}$ (solid) and $(\chi^{A})^{-1}$ (dashed) that arises from application of an adiabatic kernel $f\Hxc^{A}$ as indicated. The dotted lines show the true inverse kernel $\chi^{-1}$ (from Eq.~\ref{eq:imchi} with $q=1$, see text)}
\label{fig:adia_doub}
\end{figure}

For a resonance arising from a double-excitation however, the
analogous figures clearly show that frequency-dependence in the kernel
is absolutely required.  The KS response function of
Eq.~\ref{eq:chis_sing} now consists solely of the gently decaying first term,
$\chi\s^{\rm c}$, (left-hand panel of Figure~\ref{fig:adia_doub}) i.e. there is no bound-state contribution to
$\chi\s$ because a double-excitation has zero oscillator
strength~\cite{MZCB04}.  The true response function however displays a resonance with an antisymmetric imaginary part ($q=\pm 1$) in order to preserve the oscillator strength sum-rule (see Sec.~\ref{sec:Fano}). 
On the right panel of
Figure~\ref{fig:adia_doub} we invert this smooth $\chi\s$ to obtain
the featureless $\chi\s^{-1}$ shown (solid lines). The dotted curves represent the true $\chi^{-1}$, computed using Eq.~(\ref{eq:imchi}) for $q=1$.  {\it A
frequency-independent kernel again only shifts the real part of $\chi\s^{-1}$
uniformly as shown, but in this case it cannot generate the resonance-structure of the
true $\chi^{-1}$}, shown as dotted lines in the figure. For that, a
frequency-dependent kernel of the form derived in the next section is required. The
left-hand panel shows the real and imaginary parts of $\chi\s$ and
$\chi^{A}$, lacking resonance.

\section{Dressed TDDFT for Resonances Arising from Double Excitations}
\label{sec:dressed}
We are interested in the frequency-range near a resonance that arises
when the energy of a bound double-excitation lies above the
single-ionization threshold.  As argued in the previous section, the
KS response function displays no resonance, and an adiabatic kernel
cannot generate one. In this section we derive the form of the kernel
that is required in order to capture this kind of resonance, based on
Fano's approach. We again stay within the assumption of weak
interaction, where the dominant coupling between the KS states near the frequencies of interest, is the resonant
coupling of the bound double-excitation with the continuum states. We again assume an isolated resonance: one discrete state coupling to one continuum.

Eq.~\ref{eq:imchi} relates the imaginary part of the true response
function to that of the KS response function via the Fano lineshape. Our job is now to use Eq.~\ref{eq:fHxc} to find the implied structure of the xc kernel. 

A first simplification is that $q(\br)^2=1$. 
This follows from the sum-rule Eq.~\ref{eq:sumrule}. Within the assumption that $\vert i\rangle$ can be approximated by the KS
ground-state, then the RHS must be zero, since $\hat T$ is a one-body operator
and $\Phi_{\rm b}$ differs from the KS ground-state by two orbitals. 
That $q^2 =1$ is consistent with the oscillator strength sum-rule~(eg. Ref. \cite{Friedrich}):
both 
the KS system and the interacting system satisfy the
oscillator strength sum-rule. As the KS spectrum does not contain the
resonance arising from a doubly-excited state, this suggests that the
integrated area under the Fano lineshape factor should be zero.  That
is, the line-shape should be antisymmetric, so $q(\br)$ should be
$\pm 1$.

Examination of Eq~\ref{eq:q} reveals that $q$ must be positive for our
case: the first term in the numerator on the right is zero by the
above argument, while the second term divided by the denominator is
positive~\cite{F61}, because $\langle \Phi_{E'} \vert T\vert i\rangle$
grows larger than $\langle \Phi_{E} \vert T\vert i\rangle$ where $E-E'
>0$.

So we may conclude for the case of a double-excitation resonance, in
the weak-interaction limit, that the imaginary part of the response
function is given by:
\ben
\Im\chi(\omega) = \frac{(\omega - \omega_r +\Gamma/2)^2}{(\omega - \omega_r)^2 + (\Gamma/2)^2}\Im\chi\s(\omega)
\label{eq:imchi_doub}
\een
where $\Gamma$ and $\omega_r$ are given by Eqs.~(\ref{eq:gamma}) and~(\ref{eq:wr}), with $V_{\rm cpl}$ given by Eq.~(\ref{eq:Vcpl}). An immediate implication is that the spatial-dependence is unchanged: this is true only within the approximations stated above.

Returning to our search for $f\Hxc$, Eq.~\ref{eq:fHxc} requires that we
invert the response function.  For this, we need to first calculate
its real part, which we may obtain from the principle-value integral, Eq.~(\ref{eq:real}).
Subtracting out $\Im\chi\s$ from $\Im\chi$, we write
\ben
\Re\chi(\omega) = \Re\chi\s(\omega) + \frac{2\Gamma}{\pi}P\int_0^{\infty} \frac{\omega'(\omega' - \omega_r)\Im\chi\s(\omega')}{(\omega'^2 - \omega^2)((\omega - \omega_r)^2 +(\Gamma/2)^2)}d\omega'
\een
In the spirit of the Fano analysis, we assume that the KS orbitals are slowly-enough varying in frequency, that $\Im\chi\s$ can be pulled out of the principle-value integral. That is, that $\Im\chi\s$ is relatively flat in the region of the resonance; further away from the resonance the integral vanishes as the lineshape factor decays rapidly. 
We obtain
\begin{widetext}
\ben
\Re\chi(\omega)= \Re\chi\s(\omega) + \left(\frac{\left(\frac{\Gamma^2}{2}(\Gamma/2)^2 +(\omega^2 +\omega_r^2)\right)(1+\frac{2}{\pi}\tan^{-1}(2\omega_r/\Gamma))- \frac{\Gamma}{2\pi}\omega_r(\omega^2 - \omega_r^2 -(\Gamma/2)^2)\ln\left(\frac{\omega^2/4}{\omega_r^2 + (\Gamma/2)^2}\right)}{\left((\omega - \omega_r)^2 + (\frac{\Gamma}{2})^2\right)\left((\omega + \omega_r)^2 + (\frac{\Gamma}{2})^2\right)}\right)\Im\chi\s(\omega)
\een
\end{widetext}
Consistent with our assumptions, we take $\Gamma<< (\omega_r - I)$, and considering frequencies $\omega$ near $\omega_r$, we obtain, after some algebra,
\ben
\Re\chi(\omega) =\Re\chi\s(\omega) + \frac{\Gamma^2/2}{(\omega - \omega_r)^2 + (\frac{\Gamma}{2})^2}\Im\chi\s(\omega)
\label{eq:rechi_doub}
\een

Putting Eqs.~\ref{eq:rechi_doub} and \ref{eq:imchi_doub} together, we have
\bea
\chi = \chi\s + \frac{\Gamma(\Gamma/2 + i(\omega - \omega_r))}{(\omega - \omega_r)^2 + (\frac{\Gamma}{2})^2}\Im\chi\s
\label{eq:chi}
\eea
The complex lineshape on the right relates the interacting response function to the non-interacting one with a  dramatic resonance structure. The cross-section obtained from the imaginary part, reproduces the Fano formula.

One should not be alarmed by the poles in our
approximate $\chi$ in the upper half plane,  given that the exact $\chi$ should be
analytic in the upper half plane, and its inverse $\chi^{-1}$ analytic
for $\Im(\omega) > 0$. But our approximate kernel holds {\it only} for
real frequencies, moreover, for frequencies in the restricted range
near the resonance.

\begin{figure}
\centering
\includegraphics[height=4cm,width=7cm]{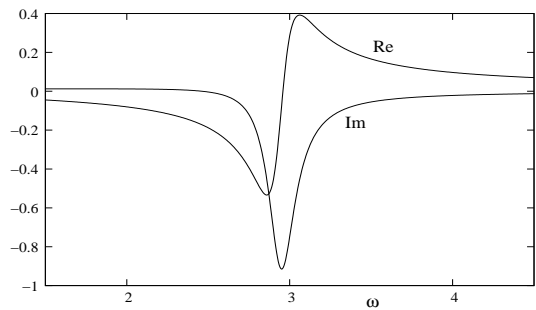}
\caption{Sketch of the Fano TDDFT kernel, Eq.~\ref{eq:fHxcfano}, that reproduces Fano's lineshape for the autoionizing resonance arising from a double-excitation.}
\label{fig:fHxc}
\end{figure}

We now use Eq.~(\ref{eq:fHxc}) to extract the frequency-dependent kernel. 
Subtracting the inverse of Eq.~\ref{eq:chi} from that of $\chi\s^{-1}$ (Eq.~\ref{eq:fHxc}) we find the Hartree-exchange-correlation kernel to be
\ben
f\Hxc(\omega) =  \chi\s^{-1}(\omega) - \left(\chi\s(\omega) + \frac{\Gamma(\Gamma/2 + i(\omega - \omega_r))}{(\omega - \omega_r)^2 + (\frac{\Gamma}{2})^2}\Im\chi\s(\omega)\right)^{-1} 
\label{eq:fHxcfano}
\een
In Eq.~\ref{eq:fHxcfano}, the spatial-dependence has been omitted for clarity, i.e. all arguments of $f\Hxc$ and $\chi\s$ are understood to be $(\br,\br',\omega)$. 
We call this the {\it Fano TDDFT kernel}, in that it reproduces exactly the Fano lineshape relative to the KS spectrum, when a bound double-excitation lies 
in the continuum. Its frequency-dependence is essential, as demonstrated by the sketches of the previous section. 
For illustration, we plot the real and imaginary parts of this in Figure~\ref{fig:fHxc}, where we take $\chi\s(\omega) \approx \frac{1}{\sqrt{\omega + I}} + \frac{i}{\sqrt{\omega - I}}$ as in the earlier plots, $\Gamma = 0.1$ and $\omega_r = 3$. 

This kernel accounts only for the coupling of the KS bound state with
the continuum, which is the dominant effect near the resonance. It becomes exact in the limit of weak interaction, for an isolated and narrow resonance, where $\Gamma$
is the smallest energy scale of the system.

We may also include this kernel on top of an adiabatic kernel, in
order to account also for mixing of non-resonant single excitations
that adiabatic TDDFT may capture well. For the case that the resonance
arising from the double-excitation is well-isolated from any other
resonances in the system, modifying Eq.~\ref{eq:chi}, we assert the {\it dressed} response-function approximation:
\ben
\chi = \chi^{A} + \frac{\Gamma(\Gamma/2 + i(\omega - \omega_r))}{(\omega - \omega_r)^2 + (\frac{\Gamma}{2})^2}\Im\chi^{A}
\label{eq:dressedachi}
\een
where $\chi^{A}(\omega)$ is the interacting response function computed
using an adiabatic xc kernel. At frequencies moving away from the resonance
arising from the double-excitation, our dressed response function
reduces to the adiabatic response function, $\chi^{A}(\omega)$; thus
our dressed response function leaves untampered  the usual reasonably
accurate response that adiabatic TDDFT gives for single-excitation
states. The corresponding dressed cross-section, obtained from its imaginary part, 
yields
\ben
\sigma(\omega) = \frac{(\omega - \omega_r +\Gamma/2)^2}{(\omega - \omega_r)^2 + (\Gamma/2)^2}\sigma^A(\omega)
\label{eq:dressedsigma}
\een
From Eq.~\ref{eq:dressedachi} we derive  the {\it dressed kernel}:
\bea
\nonumber
f\Hxc(\omega)= f\Hxc^{A} - \\
(\chi^{A})^{-1}\left(\left(1+ (\chi^{A})^{-1}\Im\chi^{A}\frac{\Gamma(\Gamma/2 + i(\omega - \omega_r))}{(\omega - \omega_r)^2 + (\frac{\Gamma}{2})^2}\right)^{-1} -1\right)
\label{eq:dressedfhxc}
\eea

In practise, computing cross-sections using our dressed TDDFT would
proceed very simply: First an adiabatic calculation would be run, as
in Refs.~\cite{SFD05,FSD03,FSD04,STFD06,STFD07,SDL95}. Then where a
bound-double excitation is known to lie (for example, by summing KS
orbital frequencies), utilize the relevant KS orbitals in
Eqs.~(\ref{eq:gamma}) and~(\ref{eq:wr}), to find the width $\Gamma$
and shift of the resonance position $F(E)$ that appears in
$\omega_r$. Then modify the cross-section computed using an adiabatic
approximation by the lineshape (Eq.~\ref{eq:dressedsigma}).  That is,
in practise, if interested in computing the cross-section, we would
not need to utilize Eq.~\ref{eq:dressedfhxc} directly, instead we would use Eq.~\ref{eq:dressedsigma}. 
Instead, Eq.~\ref{eq:dressedfhxc} and Eq.~\ref{eq:fHxcfano} are of fundamental interest here: it is the xc kernel for which approximations are needed in TDDFT, and these equations reveal the form it requires, in order to reproduce
the Fano resonance arising from a double-excitation.

\subsection{Model example}
\begin{figure}[h]
\centering
\includegraphics[height=4cm,width=8cm]{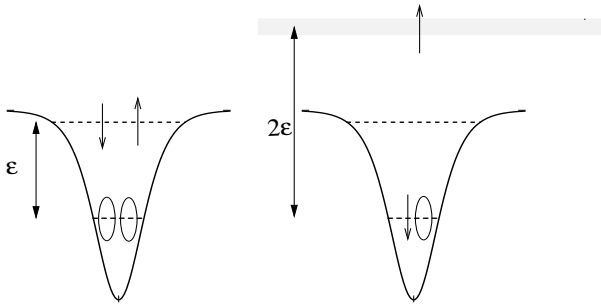}
\caption{Sketch of model potential, indicating the two excited non-interacting states, one a double-excitation that is bound (on the left) and the other a single-excitation that is unbound (on the right). Allowing the electrons to interact couples these degenerate states, creating a resonance.}
\label{fig:eckartsketch}
\end{figure}

We illustrate our results on a simple model involving two electrons in one-dimension living in an external potential of the form:
\ben
v\ext(x) = -\frac{U_0}{\cosh^2(\alpha x)} - \beta\sqrt{(1-\tanh^2(x))^3}
\label{eq:vextmodel}
\een
We choose values of parameters $U_o, \alpha$, $\beta$ such that
there are (at least) two bound single-particle states in the
non-interacting problem. A sketch is shown in
Figure~\ref{fig:eckartsketch}. A double-excitation to the first
excited orbital is shown on the left; this has energy $2\epsilon$
where $\epsilon$ is the energy difference between the single-particle
orbitals. This energy $2\epsilon$ exceeds the single-ionization
threshold for this system, and lies in the continuum; therefore the
state on the left is degenerate with a single-excitation to the
continuum, indicated on the right.
When electron-interaction is turned on, an isolated resonance is created, of the type to which our formula and analysis applies.

\begin{figure}[h]
\centering
\includegraphics[height=5cm,width=8.5cm]{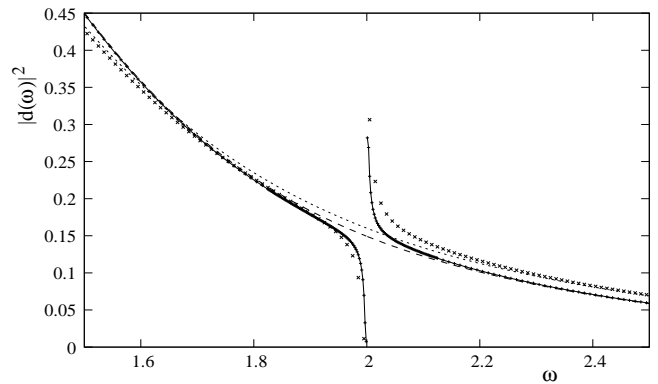}
\caption{The square of the dipole moment: Dashed line is KS, dotted line is the adiabatic TDDFT (exact-exchange), solid line is the resonance with the Fano lineshape as would be obtained with the Fano TDDFT kernel Eq.~\ref{eq:fHxcfano}, and the (x) are our dressed kernel used in conjunction with the adiabatic approximation, Eq.~\ref{eq:dressedfhxc}}.
\label{fig:eckart}
\end{figure}

We first find the KS potential and spectrum. We choose a weak delta-function interaction:
\ben
V\ee = \lambda\delta(x-x')\;,\;\; \lambda < 1.
\een
For small enough $\lambda$, exchange dominates over correlation. We use the exact-exchange approximation for two electrons:
\ben
v\xc = -v\H/2 = -\lambda n(x)/2
\een
The KS potential follows as:
\ben
v\s(x) =  v\ext +v\H + v\xc = -\frac{U_0}{\cosh^2(\alpha x)}
\een
if we take $\beta = \lambda$ in Eq.~\ref{eq:vextmodel}. 
The exact one-electron eigenstates and energies of $v\s$ (an ``Eckart well'') can be found in many quantum mechanics textbooks.
The parameters $U_0, \alpha$ are chosen such that
there are at least two bound one-particle states: we chose $U_0 = 1.875$, and $\alpha=1$, which places the non-interacting bound orbital energies at $\epsilon_0 = -1.125$ and $\epsilon_1 = -0.125$. 
 From the one-electron orbitals we calculate the KS response function $\chi\s$ and dipole moment, shown as the dashed line in Figure~\ref{fig:eckart}, after placing two electrons in the lowest orbital. As expected, the dipole moment to the continuum states is smooth and gently decaying. 

We now consider the TDDFT spectra, and will compute everything only to first-order in the interaction strength, $\lambda$, whose numerical value we take as $0.2$ in the calculations. 
 We first apply an adiabatic kernel, and again choose exact-exchange for this:
\ben
f\xc^{A}(x,x') = f\x^{A}(x,x') = -\lambda\delta(x-x')/2
\een
For the cases where the states are mixtures of single excitations,
exchange effects dominate in the weak-interaction limit, being of $O(\lambda)$ while correlation is $O(\lambda^2)$; so for small
values of $\lambda$, $f\x^{A}$ is expected to be quite accurate for
energies and oscillator strengths of states of single-excitation
character. Then
\ben
\chi^A(\br,\br', \omega) = \chi\s(\br,\br', \omega) + \frac{\lambda}{2}\int \chi\s(\br,\br_1,\omega)\chi(\br_1,\br', \omega) d^3r_1 
\een
where, on the right, we may replace $\chi$ with $\chi\s$ to get the
$O(\lambda)$ adiabatic TDDFT spectrum.
In one-dimension, from Eqs.~\ref{eq:crosssection} and~\ref{eq:sigma_sos}, 
\ben
\vert d(\omega)\vert^2 = -\frac{1}{\pi}\int x\Im\chi(x,x',\omega)x' dxdx'
\label{eq:1ddip}
\een
The KS version of Eq.~\ref{eq:imchi_ctm} simplifies to 
\ben
\Im\chi\s(x,x',\omega) = -\pi\phi_0(x)\phi_0(x')\phi_{\epsilon = \omega +\epsilon_0}(x)\phi_{\epsilon = \omega +\epsilon_0}(x')
\label{eq:imchis_ctm}
\een
Using Eq.~\ref{eq:imchis_ctm} and its principal-value integral to get the real part, Eq.~\ref{eq:1ddip} finally gives
\bea
\nonumber
\vert d^A(\omega)\vert^2= \vert d\s(\omega)\vert^2 + 2\lambda d\s(\omega)\times\\
\int dx_1\phi_0^2(x_1)\phi_\epsilon(x_1) P\int d\s(\epsilon')\phi_{\epsilon'}(x_1)\frac{\epsilon' -\epsilon_0}{\omega^2 - (\epsilon' -\epsilon_0)^2} d\epsilon'
\eea
This adiabatic dipole moment is the dotted line in  Fig~\ref{fig:eckart}: it smoothly shifts
the KS spectrum, redistributing oscillator strength from the lower to higher frequencies of the KS spectrum, but is a small and smooth correction. 

Applying now the frequency-dependent kernel Eq.~\ref{eq:fHxcfano} reveals the resonance (solid line), while the dressed kernel Eq.~\ref{eq:dressedfhxc} (x points) incorporates corrections from the adiabatic approximation as well as capturing the resonance. As mentioned in the previous section, in practise, we utilize Eq.~\ref{eq:dressedsigma}, evaluating $\Gamma$ and $\omega_r$ using the KS orbitals.



\section{Conclusions and Outlook}
We have discussed the implications of Fano's resonance
formula for the xc kernel of TDDFT. In the case
of a narrow isolated resonance involving one discrete state and one
continuum, and weak interaction, we derived the Fano-equivalent
formula for the imaginary part of the TDDFT response function
(Eq.~\ref{eq:imchi}). We illustrated how a frequency-independent
kernel applied to a largely frequency-independent KS response
function, actually yields dramatic frequency-dependent resonant
structures for the case of a bound single-excitation, but yields no
structure in the case of a bound double-excitation.

Within our assumptions, we derived the exact form of the
frequency-dependent kernel that is needed for the latter case: we call
this the Fano TDDFT kernel (Eq.~\ref{eq:fHxcfano}), in the sense that it exactly reproduces
Fano's lineshapes, when applied to the KS density-response
function. We then asserted a dressed frequency-dependent kernel (Eq.~\ref{eq:dressedfhxc}), that
accounts for the Fano effect on top of an adiabatic approximation. The
form of these kernels is of fundamental interest for TDDFT. In
practise, we propose one computes the resonance width $\Gamma$ and
position $\omega_r$ using the appropriate KS orbitals; compose from
them the Fano lineshape (Eq.~\ref{eq:dressedsigma}) and thereby modify
the adiabatic spectrum. How this will work in practise for molecules
of interest, remains to be tested.

 In this paper we considered the simplest case of a resonance arising
 from a bound double excitation: that is, when the resonance is
 isolated from all others and  involves only one continuum and one discrete
 state. We have not considered the interaction of
 resonances~\cite{CL88} that arises in most systems, nor the
 computation of individual branching ratios when several channels are
 involved~\cite{S77}. The present work is only a first
 step in uncovering how the exact xc kernel captures
 resonances in the general case, and how to approximate it in
 practise.

We gratefully acknowledge financial support from the National Science Foundation NSF CHE-0547913, and a Research Corporation Cottrell Scholar Award.

\end{document}